\documentclass[aip,cha,onecolumn,superscriptaddress,preprintnumbers,showpacs, amsmath,amssymb]{revtex4-1}
\usepackage{bm}
\usepackage[colorlinks=true,linkcolor=blue,citecolor=blue]{hyperref}
\usepackage{amsmath}
\usepackage{amssymb}
\usepackage{amsthm}
\usepackage{amsfonts}
\usepackage{enumerate}
\usepackage{latexsym}
\usepackage{ifpdf}
\usepackage{graphicx}
\usepackage{makeidx}
\expandafter\ifx\csname package@font\endcsname\relax\else
 \expandafter\expandafter
 \expandafter\usepackage
 \expandafter\expandafter
 \expandafter{\csname package@font\endcsname}
 \fi
\usepackage{color}

\begin{document}

\title{Epitaxial Growth of (1 1 1)-Oriented Spinel CoCr$_2$O$_4$/Al$_2$O$_3$ Heterostructures}

\author{Xiaoran Liu}
\email{xxl030@email.uark.edu}
\affiliation{Department of Physics, University of Arkansas, Fayetteville, Arkansas 72701, USA}

\author{D. Choudhury}
\affiliation{Department of Physics, University of Arkansas, Fayetteville, Arkansas 72701, USA}

\author{Yanwei Cao}
\affiliation{Department of Physics, University of Arkansas, Fayetteville, Arkansas 72701, USA}

\author{S. Middey}
\affiliation{Department of Physics, University of Arkansas, Fayetteville, Arkansas 72701, USA}

\author{M. Kareev}
\affiliation{Department of Physics, University of Arkansas, Fayetteville, Arkansas 72701, USA}

\author{D. Meyers}
\affiliation{Department of Physics, University of Arkansas, Fayetteville, Arkansas 72701, USA}

\author{J.-W. Kim}
\affiliation{X-ray Science Division, Argonne National Laboratory, Argonne, Illinois 60439, USA}

\author{P. Ryan}
\affiliation{X-ray Science Division, Argonne National Laboratory, Argonne, Illinois 60439, USA}

\author{J. Chakhalian}
\affiliation{Department of Physics, University of Arkansas, Fayetteville, Arkansas 72701, USA}

\begin{abstract}

High quality (1 1 1)-oriented CoCr$_2$O$_4$/Al$_2$O$_3$ heterostructures were synthesized on the sapphire (0 0 0 1) single crystal substrates in a layer-by-layer mode. The structural properties are demonstrated by $in$-$situ$ reflection high energy electron diffraction, atomic force microscopy, X-ray reflectivity and X-ray diffraction. X-ray photoemission spectroscopy confirms that the films possess the proper chemical stoichiometry. This work offers a pathway to fabricating spinel type artificial quasi-two-dimensional frustrated lattices by means of geometrical engineering. 

\end{abstract}

\maketitle

\newpage

Recently, a great number of correlated complex heterostructures have been proposed to possess exotic quantum phenomena such as quantum Hall effect\cite{Di Xiao}, topological insulator\cite{Ruegg, Hu, Kargarian,A.Ruegg} and quantum anomalous Hall effect\cite{Wang}. Among these artificial systems, there exists a common material design theme regarded as geometrical lattice engineering, namely growing a two dimensional {\textquotedblleft}active material{\textquotedblright} sandwiched by wide band-gap insulators along the high index crystallographic [1 1 1] directions.\cite{A.Ruegg} In such a way, the lattice geometries, electronic bandwidth, chemical environments and electron hopping routes of the materials are tuned to be tremendously different from the bulk. As a result, collective emergent quantum states are expected to emerge. Towards this goal, two main challenges hinder the fabrications of high quality (1 1 1) oriented heterostructures. First, the atomic layers along such direction are not charge neutral planes, the strong polarity mismatch between the film and the substrate  may lead to the interfacial reconstruction, defect and impurity phase formation\cite{Nakagawa,Blok,Sri_srp}. Second, due to the limited availability of (1 1 1) oriented substrates, large lattice mismatch (5\% and more) offen prevents forming flat and homogenious films. To date, most of the pioneering experimental efforts have focused on fabricating artificial perovskite (ABO$_3$) type [1 1 1] superlattices\cite{Sri, Gibert, Herranz, Sri_arkive} due to the relatively simple lattice geometries. However there are few works published on the growth of other types of two dimensional correlated superlattices such as spinel type (AB$_2$O$_4$) which constitutes a family with rich and intriguing physical properties but crystallizes in a far more complex structure.\cite{Jak_nm}
     
CoCr$_2$O$_4$ (CCO) has a prototypical normal spinel structure in which all the Co$^{2+}$ ions are tetrahedrally coordinated by four oxygen ions and the Cr$^{3+}$ ions are octahedrally coordinated by six oxygen ions\cite{Menyuk} [see Fig. 1~(a)]. This material exhibits several interesting physical phenomena including ferrimagnetism\cite{Singh}, mutiferroicity\cite{Yamasaki}, and magnetic frustration\cite{Tomiyasu}. The crystal structure of CCO viewed along the (1 1 1) orientation is composed of alternative ionic planes. The O sublattice has a dimension of 2.82 {\AA} of the surface unit cell and presents a cubic close packing (CCP) arrangement of atoms in the form of ABCABC. The Co and Cr ionic planes between the O layers have three types of geometrically frustrated structures: triangular Co plane (T plane), triangular Cr plane (T' plane) and kagom\'{e} Cr plane (K plane) as indicated in Fig. 1~(b). In the bulk 18 ionic planes form a full repeat pattern in the form of [O-K-O-T-T'-T]$_3$ \cite{Vaz} with the basic repeat unit [O-K-O-T-T'-T] as shown in Fig. 1~(b). In what follows this repeat unit is defined as {\textquotedblleft}1CCO{\textquotedblright}. 

In this Letter, we report on the fabrication of a new heterostructure CoCr$_2$O$_4$ (CCO)/Al$_2$O$_3$ (AlO) along the [1 1 1] direction. $\alpha$-AlO is selected as the spacer because its O sublattice consists of the hexagonal close packing (HCP) with a dimension of 2.74 {\AA}, which is very close to that of CCO.\cite{Cousins,Kirfel} The compatibility of CCP and HCP lattices offers the perfect structural and chemical continuity between CCO and AlO layers, which effectively eliminates the polar and lattice mismatch issues. Good crystallinity and smooth surface morphology were demonstrated by $in$-$situ$ high pressure reflection-high-energy-electron-diffraction (RHEED) and atomic force microscopy (AFM). Synchrotron based X-ray diffraction (XRD) scan elucidated the growth direction of the heterostructures was along [1 1 1] with no impurity phases. Detailed film thickness of each layer and interfacial roughness were investigated by X-ray reflectivity (XRR). The chemical stoichiometry and core level electronic structures of CCO layer was investigated by monochromatic X-ray photoemission spectroscopy (XPS) indicating the oxidation states of Co to be 2+ and Cr to be 3+. The combined results allow to establish the high quality of the synthesized heterostructures with the quasi-two-dimensional frustrated kagom\'{e} and triangular layers.           

A series of 2CCO/mAlO (m = 1 and 2 unit cells) superlattices was fabricated by the pulsed laser interval deposition method\cite{Kareev} using a KrF excimer laser operating at $\lambda$  $=$ 248 nm. The substrate was maintained at 800 $^{\circ}$C under a partial pressure of 1 - 5 mTorr of oxygen during the deposition with the laser's intensity $\sim$ 2 J/cm$^2$ at the target. The ablation frequencies for CCO and AlO were 4 Hz and 2 Hz, respectively. Samples were annealed at the growth condition for 15 min and then cooled down to room temperature. The entire growth process was monitored by $in$-$situ$ RHEED with the incident electron beam along the [1 $\bar{1}$ 0 0] direction of the substrate. 1 $\mu$m by 1 $\mu$m AFM images were scanned over several different areas of the sample surface after the post anneal process to check the surface morphology and flatness. Detailed information about the thickness of each layer and interfacial roughness was obtained from the fittings of the X-ray reflectivity (Cu K$_{\alpha1}$ line ($\lambda$ $=$ 1.5406 {\AA})) data. Synchrotron based X-ray diffraction measurement was performed at the beamline 6-ID-B of Argonne National Laboratory. The chemical valences and proper stoichiometry of the samples was investigated by core-level XPS measurements with monochromatic Al K$_{\alpha}$ source.   

First we discuss the structural properties of the heterostructures. Fig. 1~(c)-(f) RHEED patterns present a detailed overlook of the deposition process. The clear specular and off-specular reflections on the zero-order Laue circle together with the Kikuchi lines in Fig. 1~(c) confirm a flat and good crystalline substrate surface. Once the CCO layer is formed on the substrate, two additional streaks occur in the middle of the specular and off-specular spots as shown in Fig. 1~(d). These half order reflections are the characteristic signature of spinel thin films grown on sapphire substrate and have also been observed in other works.\cite{Xiaoran, Y.Gao, Matzen} The evolution of these RHEED images confirms that the epitaxial growth relation is CCO (1 1 1)[1 $\bar{1}$ 0] $|$$|$ $\alpha$-AlO (0 0 0 1)[1 $\bar{1}$ 0 0]. When the first AlO is deposited onto the CCO layers, the typical spinel half orders can be still observed but become weaker [Fig. 1~(e)]. This is likely due to the formation of the $\gamma$-AlO phase which has an Al-deficient spinel structure\cite{Siqi}. Since the $\gamma$-AlO has lower surface energy\cite{McHale} and more flexibility in cation distribution\cite{Siqi}, it is likely to appear at the initial growth stage.\cite{Siqi, Tanner} As the second AlO layer is deposited, however, the half order streaks vanish and the RHEED pattern looks closer to the substrate pattern, as displayed in Fig. 1~(f). This observation indicates that the second AlO layer evolves into the $\alpha$-AlO phase. Further AlO deposition proudces the same RHEED patterns as Fig. 1~(f) which reveals the $\alpha$-AlO phase attains stability after the formation of the first transitional $\gamma$-AlO phase. The post-annealed RHEED images are shown in Fig. 1~(g). As seen, The distinct reflection streaks are maintained throughout the entire deposition implying the well developed crystallinity and the two dimensional layer-by-layer growth mode. To further examine the surface lattice symmetry, the samples were rotated by 60$^{\circ}$ steps relative to the original position. A characteristic RHEED image shown in Fig. 1~(h) reveal the same pattern, which confirms that the surface lattice has indeed a six-fold symmetry. In addition, the average rms surface roughness of the resultant samples is about 2 {\AA} (not shown), which provides another strong evidence of the smooth surface morphology.             
 
More detailed information about the thickness, the surface and interface roughness of these sets of superlattices was investigated by XRR. Two typical reflectivity curves with simulations for a thin sample [2CCO/1AlO]$_4$ and a thick [2CCO/2AlO]$_{10}$ sample are shown in Fig. 2~(a). The distinct intensity oscillations ($\it{i.e.}$ Kiessig fringes) over the whole scan range indicate abrupt interfaces and flat surface crystallinity of the samples. All of the fit parameters and the corresponding values are given in Table I. As seen, the fitted thickness of each layer is in excellent agreement with the values estimated from the specific number of laser pulses and RHEED oscillations of the reflected intensity. Furthermore, both superlattices maintain low roughness throughout the whole sample which corroborates the two-dimensional layer-by-layer stackings of each part also  monitored by RHEED. The overall structural quality of the films have been investigated by synchrotron based XRD. The diffraction scan in the vicinity of sapphire (0 0 6) reflection on the [2CCO/1AlO]$_4$ is shown in Fig. 2~(c). As seen, besides the sharp substrate (0 0 6) peak, the broad film peak surrounded by thickness fringes on both sides which correspond to the (2 2 2) reflection of the superlattice is also clearly observed. Based on the XRD data, good crystallinity and proper (1 1 1) orientation of the heterostructures has been confirmed.       
 
Next we turn our attention to the electronic properties of these samples. In order to investigate the valences and chemical stoichiometry of the sandwiched CCO layers, core-level XPS measurements were performed on the superlattices. Fig. 3 shows the core-level Co 2$\it{p}$ and Cr 2$\it{p}$ spectra collected on the representative 2CCO/1AlO sample. As seen on the left panel, the Co 2$\it{p}$ spectrum consists of the spin-orbit split peaks corresponding to Co 2$\it{p}$$_{3/2}$ and Co 2$\it{p}$$_{1/2}$ at binding energies of 781.66 eV and 797.25 eV, respectively, and has an energy separation denoted as $\Delta$$_{Co, 2p}$ $\sim$15.6 eV between them. The observed binding energy separation of 15.6 eV is characteristic for compounds containing Co$^{2+}$ ions in a tetrahedral oxide environment\cite{Dillard, Kumar}. We also note that the binding energy separation in case of Co$^{3+}$ containing compounds is usually smaller ($\sim$ 15.1 eV) \cite{Okamoto}. 

In addition to the spin-orbit split main peaks, the Co 2$\it{p}$ spectrum also consists of two intense shake-up satellite peaks located on the higher energy sides of the Co 2$\it{p}$$_{3/2}$ and Co 2$\it{p}$$_{1/2}$ peaks, which are highlighted by the red arrows. It is well known\cite{Okamoto,Bocquet} that while Co$^{2+}$ compounds show intense shake-up satellite peaks in Co 2$\it{p}$ spectrum, the same features are much weaker in case of Co$^{3+}$ containing compounds. In case of Co oxide samples, the shake-up satellite corresponds to the charge transfer excitation from a ligand oxygen 2$\it{p}$ level to empty Co 3$\it{d}$ states, and hence the relative energy separation from the shake-up satellite to its corresponding main peak is a distinct signature of the Co charge state. As seen in Fig. 3, the separation between the Co 2$\it{p}$$_{3/2}$ peak and the corresponding shake-up satellite is 4.6 eV and is very close to that observed in other Co$^{2+}$ compounds.\cite{Dillard} On the right panel, the Cr 2$\it{p}$ spectrum similarly consists of the spin-orbit split Cr 2$\it{p}$$_{3/2}$ and Cr 2$\it{p}$$_{1/2}$ peaks at the binding energies of 577.15 eV and 587.20 eV, respectively. The binding energy difference of $\sim$10.0 eV between the Cr 2$\it{p}$$_{3/2}$ and Cr 2$\it{p}$$_{1/2}$ peaks agrees very well with other Cr$^{3+}$ containing compounds\cite{Kumar}. Based on these XPS results, the valences of Co and Cr in our superlattice samples are assured to be 2+ and 3+, respectively. 

In summary, we have fabricated artificial high quality (1 1 1)-oriented CCO/AlO superlattices with geometrically frustrated lattices on sapphire (0 0 0 1) substrate. The comprehensive structural and electronic characterizations established the layer-by-layer growth with proper chemical stoichiometry which consist of alternative stacking of kagom\'{e} and triangular atomic planes. Both the CCO and the AlO thickness can be digitally controlled. The presented results pave a way for fabricating unique two dimensional transition metal heterostructures on which a plethora of intriguing quantum phenomena can be expected.

The authors deeply acknowledges numerous fruitful discussions with  D. Khomskii and G. Fiete. J.C. was supported by the DOD-ARO under Grant No. 0402-17291 and in part by the Gordon and Betty Moore Foundation. Work at the Advanced Photon Source, Argonne is supported by the U.S. DOE under Grant No. DEAC02Â06CH11357.

\begin{table}[t]
\caption{Reflectivity Fit Parameters}
\centering
\setlength{\tabcolsep}{10pt}
\begin{tabular}{l	 l l l l l l}
\hline\hline
Sample & Items & Top AlO (\AA) & Superlattice (\AA) & Bottom CCO (\AA) \\ [0.5ex]
\hline
& thickness & 14.6 & 9.9/13.6 & 11.0 \\[-1ex]
\raisebox{1.5ex}{[2CCO/1AlO]$_4$} & roughness & 3.5 & 2.6 & 1.0 \\[1ex]
& thickness & 30.8 & 9.8/27.0 & 12.9 \\[-1ex]
\raisebox{1.5ex}{[2CCO/2AlO]$_{10}$} & roughness & 5.0 & 3.4 & 0.7 \\[1ex]
\hline\hline
\end{tabular}
\label{Table I}
\end{table}
\clearpage

\newpage
\begin{figure}[t]\vspace{-0pt}
\includegraphics[width=1\textwidth]{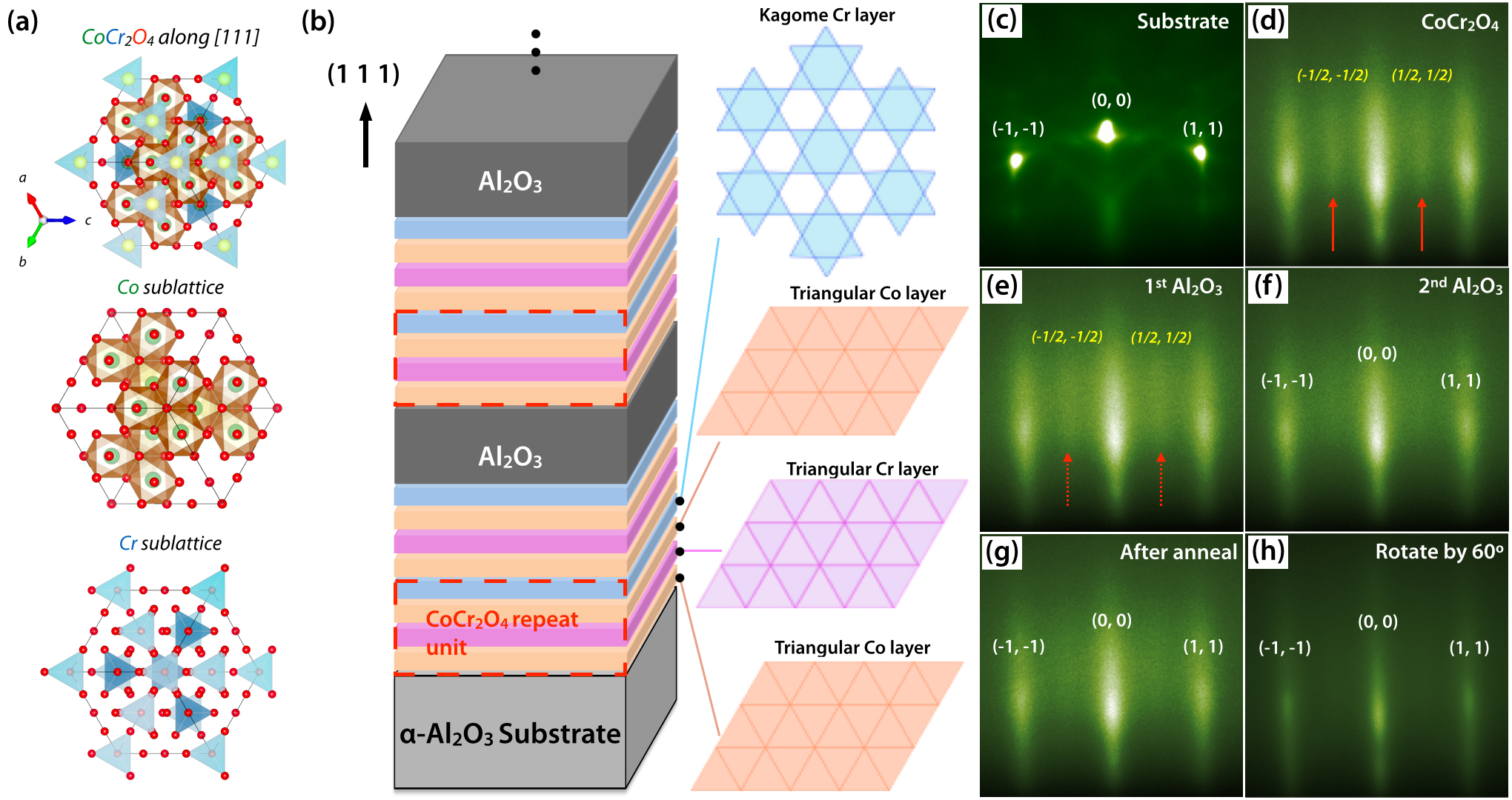}
\caption{\label{} (a) CCO bulk unit cell viewed along the (1 1 1) orientation. The Co and Cr sublattices are shown respectively. (b) Schematics of the 2CCO/mAlO heterostructures. The CCO repeat unit along the (1 1 1) orientation is emphasized by the red dashed lines. The lattice structures of each ionic planes in one repeat unit are depicted in different colors (blue, orange and pink). Note, only the Co and Cr ionic planes are shown in the picture for clarity. (c)-(g) RHEED patterns during the deposition process. The observed half order reflections are  pointed out by the red arrows and coordinated in yellow in the pictures. Note, the incident electron beam of the RHEED is fixed along the [1 $\bar{1}$ 0 0] direction of the substrate. (h) Characteristic RHEED images corresponding to (g) collected after rotating the film by 60$^{\circ}$ steps.}
\end{figure}

\newpage
\begin{figure}[t]\vspace{-0pt}
\includegraphics[width=1\textwidth]{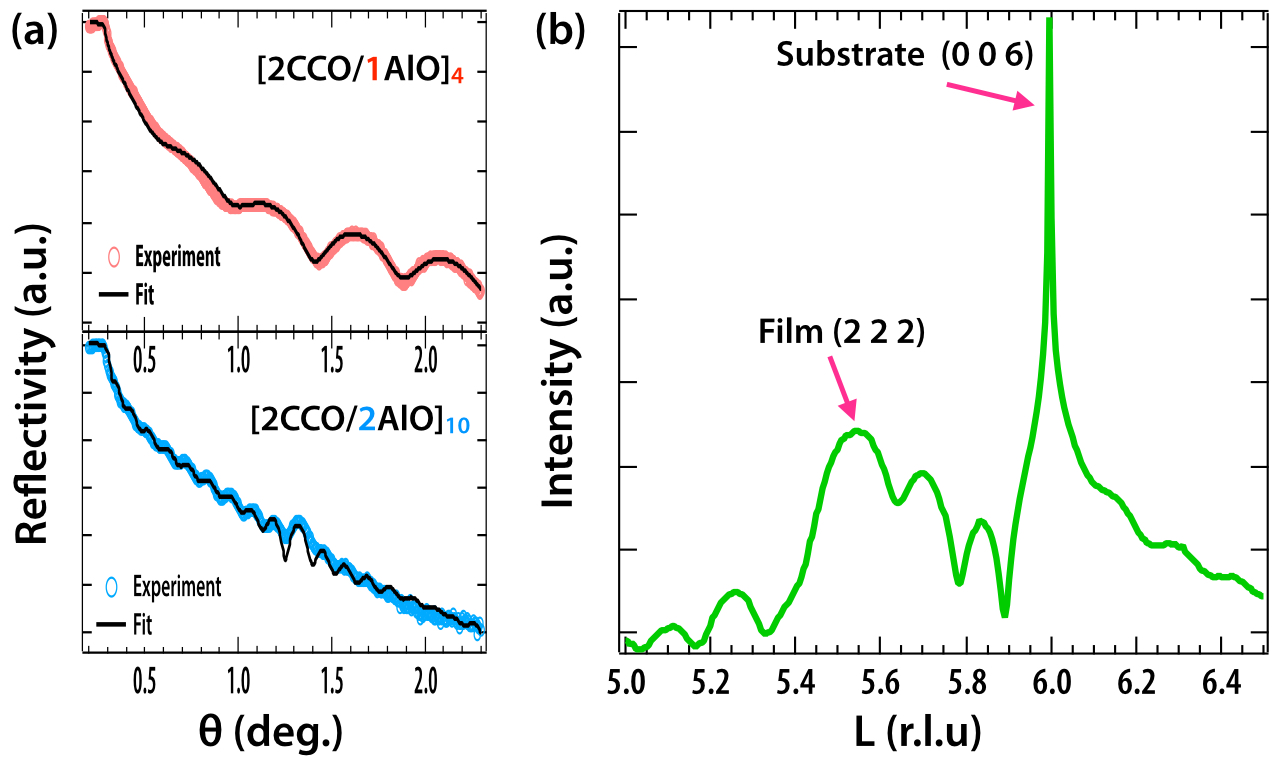}
\caption{\label{}(a) XRR data of the thin [2CCO/1AlO]$_4$ (top panel) and the thick [2CCO/2AlO]$_{10}$ (bottom panel) samples, respectively. The solid black curves are the simulated data. (b) XRD scan of the [2CCO/1AlO]$_4$ sample around the (0 0 6) reflection of the substrate. The film and the substrate peaks are both labeled on the graph. The incident wavelength is $\lambda$  $=$ 1.4932 {\AA}.}
\end{figure}

\newpage
\begin{figure}[t]\vspace{-0pt}
\includegraphics[width=1\textwidth]{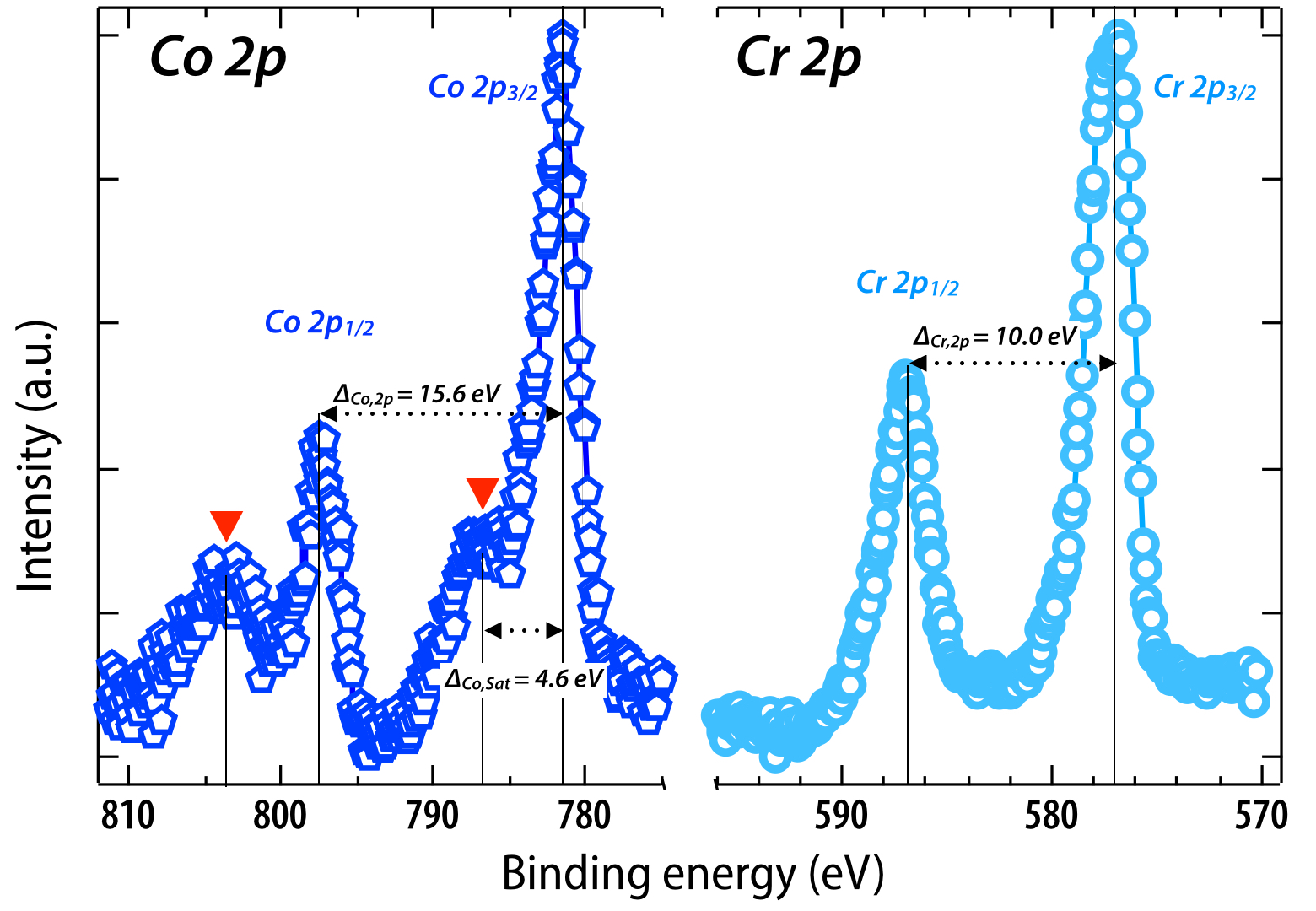}
\caption{\label{}Core-level XPS data of the Co and the Cr 2$\it{p}$ levels of the samples. The red arrows represent the Co$^{2+}$ shake-up satellites described in the text. The binding-energy separations of the Co 2$\it{p}$ peaks, Cr 2$\it{p}$ peaks and the energy difference between Co 2$\it{p}$$_{3/2}$ and its corresponding satellite peak, are denoted as $\Delta$$_{Co, 2p}$, $\Delta$$_{Cr, 2p}$, and $\Delta$$_{Co, Sat}$, respectively on the graph.}
\end{figure}

\end{document}